\documentclass[twocolumn,pre]{revtex4}

\pdfoutput=1

\newcommand{\ave}[1]{\left \langle #1 \right \rangle}
\newcommand\Ro[1][\relax]{\ifx\relax#1 \ensuremath{\mathcal{R}_0}
  \else \ensuremath{\mathcal{R}_{0,#1}} \fi}
\usepackage{amsmath}
\usepackage{graphicx}

\begin{document}
\title{Percolation in clustered networks}
\author{Joel C. Miller}
\email{joel.c.miller.research@gmail.com}
\affiliation{Fogarty Institute and Harvard School of Public Health}
\date{\today}

\begin{abstract}
  The social networks that infectious diseases spread along are
  typically clustered.  Because of the close relation between
  percolation and epidemic spread, the behavior of percolation in such
  networks gives insight into infectious disease dynamics.  A number
  of authors have studied clustered networks, but the networks often
  contain preferential mixing between high degree nodes.  We introduce
  a class of random clustered networks and another class of random
  unclustered networks with the same preferential mixing.  We
  analytically show that percolation in the clustered networks reduces
  the component sizes and increases the epidemic threshold compared to
  the unclustered networks.
\end{abstract}

\maketitle
Classical random networks contain few short cycles, and the proportion
of nodes in short cycles goes to zero as the number of nodes
increases.  In contrast social networks typically contain many short
cycles.  We refer to such networks as \emph{clustered} networks.  The
impact of clustering on percolation properties is usually difficult to
calculate because cycles prevent the use of branching process
arguments, but it is widely expected that clustering significantly
alters percolation.

Typically studies of infectious disease spread assume that outbreaks
begin with a single infected node.  The disease travels to each
susceptible neighbor independently with probability $T$, the
\emph{transmissibility}, and the node recovers.  The process repeats.
We focus on diseases for which recovery provides immunity, so
recovered nodes are not susceptible.  Typically the outbreak dies out
stochastically or becomes an epidemic and spreads until the number of
susceptible nodes is reduced.

It is well-established that for fixed $T$, the epidemic spread can be
mapped to a bond percolation problem wherein each edge is kept with
probability
$T$~\cite{newman:spread,grassberger:percolation,cardy:percolation,miller:heterogeneity,kenah:networkbased,ludwig}.
If we perform percolation on the network and then choose the initial
infection, the disease spreads from that initial infection along edges
of the percolated network, and so an epidemic occurs iff the initial
node is in the giant component. The size of the epidemic matches the
size of the giant component.  This establishes that the probability
and fraction infected in epidemics are equal if $T$ is fixed and all
edges are independent~\footnote{Care must be taken that no dependence
  between edges arises.  Such a dependence can arise from, for
  example, heterogeneity in duration of
  infection~\cite{miller:heterogeneity,kenah:second}.}.

Because social networks frequently exhibit clustering, a number of
studies have investigated the impact of clustering on epidemic
problems~\cite{miller:RSIcluster,newman:clustering,serrano:prl,serrano:2,britton:cluster,eames:clustered,keeling:local,bansal:thesis}.
Some have found that clustering reduces the sizes of
epidemics and raises the epidemic threshold.  That is, clustering
reduces the size of giant components and raise the percolation
threshold.  However, others have shown that clustering appears
to reduce the threshold.  Consequently epidemics should be possible at
lower transmissibility in the presence of clustering.

This discrepancy occurs because there are many ways used to generate
clustered networks, and each nework class results in different behaviors.
It is difficult to separate the impact of clustering from other
features introduced by the network generation process.

In this article we introduce a new algorithm to generate random
clustered networks~\footnote{This algorithm was simultaneously
  developed by~\cite{newman:cluster_alg}}.  The clustered networks
have correlations between degrees in a well-defined manner which can
lead to assortativity, the tendancy for nodes to contact nodes of
similar degree.  We show how to generate unclustered networks with the
same correlations.  We can make analytic comparisons between the two,
and so clearly separate the effect of clustering from degree
correlations.  We show that although the clustered networks can have a
reduced threshold compared to purely random networks of the same
degree distribution, that is entirely an artifact of the
assortativityx.  Compared to an unclustered network of the same degree
correlations, the clustered networks result in smaller epidemics and
higher epidemic threshold.

This article is organized as follows: we first introduce our clustered
and unclustered networks.  We then calculate and compare the
epidemiological quantity $\Ro$ which measures how many new infections
a typical infected node causes.  Finally, we calculate the final
size/probability of epidemics assuming constant
$T$.

\section{The Networks}
We model our approach after standard algorithms for Configuration
Model (CM)
networks~\cite{newman:structurereview,MolloyReed,bollobas:CM}.  CM
networks are useful because all edges from a node are independent of
one another, in the sense that whether an epidemic results from
following one edge is independent of the result along any other edge
because short cycles are negligible.

\subsection{Clustered Networks}
We begin with $N$ nodes.  To each node $u$ we assign two degrees, an
\emph{independent edge} degree $k_I$ and a triangle degree
$k_\triangle$.  The joint probability of $k_I$ and $k_\triangle$ is
given by $p(k_I,k_\triangle)$.  Then $u$ will be part of $k_\triangle$
triangles and have $k_I$ other edges.  Each triangle and edge from $u$
will be independent of other triangles and edges in the same way that
edges in CM networks are independent.

We create an independent stub list and a triangle stub list.  We place
$u$ into the independent stub list $k_I$ times and into the triangle
stub list $k_\triangle$ times.  Once all nodes are placed into the
lists, we randomize them.  We then take the pairs of nodes in
positions $2n$ and $2n+1$ of the independent list and join them, and
the triples in positions $3n$, $3n+1$, and $3n+2$ and join them into a
triangle.  Some repeated edges or loops or short cycles other than the
triangles we impose may appear, but their impact is negligible as $N
\to \infty$~\footnote{In essence we have created
  a generalization of an edge which corresponds to a triangle.  We
  could create other more general structures in much the same way}.

This algorithm inevitably segregates those nodes with a high
proportion of triangles from those nodes with a low proportion of
triangles.  If the degrees of nodes with many triangles
differs from the degrees of nodes with few triangles, then
this effect will cause correlation of different degrees.  In
order to understand the impact of clustering, we must be able to
compare percolation in these clustered networks with percolation in
networks whose nodes are segregated in the same way.

\subsection{Unclustered, Segregated Networks}

For comparative purposes we develop a corresponding unclustered
network with the same segregation as the clustered networks.  Given
the joint distribution $p(k_I,k_\triangle)$ of independent and
triangle degrees, we create a new network where nodes are assigned
\emph{blue} and \emph{red} degrees such that $k_b=k_I$ and $k_r =
2k_\triangle$.  The joint distribution is given by $p_u(k_b,k_r) =
p(k_b,k_r/2)$.

We proceed as before.  We create a blue and red list, and pair nodes
in positions $2n$ and $2n+1$ in the blue list and then repeat with the
red list, joining pairs, not triples.  The resulting network has the
same segregation as the corresponding clustered network, but short
cycles are negligible.

\section{$\Ro$}
$\Ro$ is usually defined as the number of new infections caused by an
average infected individual.  Occasionally alternate definitions are
used, but in some way it represents the number of new infections
attributed to an average infected individual.  $\Ro=1$ is the
threshold below which epidemics are impossible (\emph{i.e.}, the
percolated network has no giant component).  If $\Ro>1$ then epidemics
are possible, but not guaranteed.

\subsection{Clustered Networks}

To simplify the analysis, first assume that $u$, $v$, and $w$ are
members of a triangle and $u$ becomes infectious first.  There are
multiple ways that both $v$ and $w$ can become infected from edges
within the triangle, but they all have the same impact on the
epidemic.  It is convenient to treat infections of $v$
and $w$ as if they came from from $u$ regardless of the actual path
followed.  

Thus if $u$ becomes infected, then with probability $2T^2(1-T)+T^2 =
3T^2-2T^3$ it is credited with infecting both $v$ and $w$, and with
probability $T(1-T)^2$ it is credited with infecting just $1$.  With
probability $(1-T)^2$ it infects neither.  In spirit this approach is
similar to that of~\cite{becker:control}.  For book-keeping purposes,
we define the rank $s$ of a node as follows: the index case is given
rank $0$.  Each node $v$ is then assigned rank $s$ to be the shortest
path from the index case to $v$, bearing in mind the rule above for
crediting infections.

This allows us to define a $2\times 2$ next-generation
matrix~\cite{diekmann}.  We separate those nodes infected along an
independent edge from those nodes infected along a triangle
edge~\footnote{Without our simplification, we would need to further
  subdivide those infected along triangle edges into those whose other
  neighbor is still susceptible from those whose other neighbor is
  also infected}.  We define $c_{II}$ and $c_{\triangle I}$ to be the
number of infections that a node infected from an independent edge is
expected to cause along independent and triangle edges respectively.
We symmetrically define $c_{I\triangle}$ and $c_{\triangle
  \triangle}$.  If $n_I(s)$ and $n_\triangle(s)$ are the number of
nodes of rank $s$ which were infected along independent and triangle
edges respectively, then
\[
\begin{pmatrix}
n_I(s+1)\\n_\triangle(s+1)
\end{pmatrix}
=\begin{pmatrix} 
c_{II} & c_{I\triangle}\\
c_{\triangle I} & c_{\triangle \triangle}
\end{pmatrix} 
\begin{pmatrix}n_I(s)\\n_\triangle(s)
\end{pmatrix} \, ,
\]
where $c_{II} =  \frac{T\ave{K_I^2-K_I}}{\ave{K_I}}$, \ $c_{\triangle I}
= \frac{2T(1+T-T^2)\ave{K_IK_\triangle}}{\ave{K_I}}$, \
$c_{I\triangle} = \frac{T\ave{K_IK_\triangle}}{\ave{K_\triangle}}$,
and $c_{\triangle\triangle} =
\frac{2T(1+T-T^2)\ave{K_\triangle^2-K_\triangle}}{\ave{K_\triangle}}$.  

The dominant eigenvalue of this matrix is $\Ro$.  We generally want to
determine $T$ such that $\Ro<1$.  Substituting $\Ro=1$ into the
characteristic equation gives
\begin{align*}
&\left( T \frac{\ave{K_I^2-K_I}}{\ave{K_I}} -1 \right ) \left(
  2T(1+T-T^2)\frac{\ave{K_\triangle^2-K_\triangle}}{\ave{K_\triangle}}-1\right)
\\
&=
2T^2(1+T-T^2)\frac{\ave{K_IK_\triangle}^2}{\ave{K_I}\ave{K_\triangle}}
\end{align*}
The value $T=T_c$ that solves this equation is the threshold
transmissibility below which epidemics are impossible.

The original network has a giant component if $\Ro>1$ when $T=1$.
Thus the conditions for a giant component are
\[
\frac{\ave{K_I^2-K_I}}{2\ave{K_I}} +
\frac{\ave{K_\triangle^2-K_\triangle}}{\ave{K_\triangle}} > 1
\]
and/or
\[
\left(\frac{\ave{K_I^2-K_I}}{\ave{K_I}}-1 \right)
\left(2\frac{\ave{K_\triangle^2-K_\triangle}}{\ave{K_\triangle}} -
  1\right) <
2\frac{\ave{K_IK_\triangle}^2}{\ave{K_I}\ave{K_\triangle}}
\]
The only networks for which the first condition applies but not the
second are networks with enough independent edges and triangle edges
such that a giant component exists soley within the independent edges
and a giant component exists soley within the triangle edges. 

\subsection{Unclustered, Segregated Network}
We define $n_b(s)$ and $n_r(s)$ in the same manner, except that
triangles need not be considered.  Then
\[
\begin{pmatrix}
n_b(s+1)\\n_r(s+1)
\end{pmatrix}
=\begin{pmatrix} 
c_{bb} & c_{br}\\
c_{rb} & c_{rr}
\end{pmatrix} 
\begin{pmatrix}n_b\\n_r
\end{pmatrix}
\]
where $c_{bb} = \frac{T\ave{K_b^2-K_b}}{\ave{K_b}}$, \ $c_{br} =
\frac{T\ave{K_bK_r}}{\ave{K_r}}$, \ $c_{rb} =
\frac{T\ave{K_rK_b}}{\ave{K_b}}$, and $c_{rr} =
\frac{T\ave{K_r^2-K_r}}{\ave{K_r}}$.  Substituting $\Ro=1$ into the
characteristic equation finds the epidemic threshold
\[
\left( T\frac{\ave{K_b^2-K_b}}{\ave{K_b}} - 1 \right)
\left(T\frac{\ave{K_r^2-K_r}}{\ave{K_r}}-1\right) =
T^2\frac{\ave{K_rK_b}^2}{\ave{K_r}\ave{K_b}}
\]
The network has a giant component when
\[
\frac{\ave{K_b^2-K_b}}{2\ave{K_b}} + 
\frac{\ave{K_r^2-K_r}}{2\ave{K_r}}>1
\]
and/or
\[\left(\frac{\ave{K_b^2-K_b}}{\ave{K_b}} - 1 \right)
\left(\frac{\ave{K_r^2-K_r}}{\ave{K_r}}-1\right) <
\frac{\ave{K_rK_b}^2}{\ave{K_r}\ave{K_b}} \, .
\]
The difference between these conditions and those of the corresponding
clustered network comes from the fact that
$2\ave{K_\triangle^2-K_\triangle}/\ave{K_\triangle} <
\ave{K_r^2-K_r}/\ave{K_r}$.  From this it can be shown that the
epidemic threshold occurs at smaller $T$ for the unclustered network.

\section{Calculating Giant Component Size}
To calculate the fraction of nodes in the giant component, it suffices
to calculate the probability that a random node is \emph{not} part of
the giant component.  These calculations have been done for CM
networks by~\cite{miller:heterogeneity,newman:spread,kenah:second}.

\subsection{Clustered Network}
We follow the approach
of~\cite{miller:heterogeneity,miller:RSIcluster}.  A related approach
is given by~\cite{newman:spread}.  

We let $f$ be the probability a random node $u$ is not part of the giant
component.  We have
\[
f = \sum_{k_I,k_\triangle}
p(k_I,k_\triangle)g_I^{k_I}g_\triangle^{k_\triangle} \, ,
\]
where $g_I$ and $g_\triangle$ are the probabilities that an
independent edge or a triangle respectively does not connect to the
giant component.  To find $g_I$, we note that there are two ways an
edge can fail to connect $u$ to the giant component: It may be deleted
in the percolation process with probability $1-T$, or it may be kept,
but $v$, the node reached, is not part of the giant component.  We
have
\[
g_I = 1-T+Th_I
\]
where $h_I$ is the probability that a node $v$ reached along an
independent edge is not part of the giant component.  To calculate
$h_I$ we note that $v$ is selected proportional to $k_I$, but only has
$k_I-1$ susceptible neighbors along independent edges.  We get
\[
h_I = \frac{1}{\ave{K_I}} \sum_{k_I,k_\triangle} k_I
p(k_I,k_\triangle) g_I^{k_I-1}g_\triangle^{k_\triangle} \, .
\]
For $g_\triangle$ we get
\[
g_\triangle = [1-T+Th_\triangle]^2 - 2T^2(1-T)h_\triangle(1 -
h_\triangle) \, , 
\]
where $h_\triangle$ is the probability a node reached along a triangle
edge does not connect to the giant component through any edge not in
the triangle.  We find
\[
h_\triangle = \frac{1}{\ave{K_\triangle}} \sum_{k_I,k_\triangle}
k_\triangle p(k_I,k_\triangle) g_I^{k_I} g_\triangle^{k_\triangle-1}
\, .
\]

The resulting system of equations for $g_I$, $g_\triangle$, $h_I$, and
$h_\triangle$ can be solved iteratively, and the result gives $f$.

\subsection{Unclustered, Segregated Network}

To find $f_u$, the probability a random node in the unclustered
network is not part of the giant component, we proceed similarly.  We
find
\begin{align*}
f_u &= \sum_{k_r,k_b} p_u(k_b,k_r)g_b^{k_b}g_r^{k_r}\\
g_b  &= 1-T +Th_b\\
g_r &= 1-T +Th_r\\
h_b &= \frac{1}{\ave{K_b}} \sum_{k_r,k_b} k_b
p_u(k_b,k_r)g_b^{k_b-1}g_r^{k_r}\\
h_r &= \frac{1}{\ave{K_r}} \sum_{k_r,k_b} k_r
p_u(k_b,k_r)g_b^{k_b}g_r^{k_r-1} \, .
\end{align*}
It can be shown that $f_u<f$
for the equivalent degree distributions.  Consequently the size of the
giant component is smaller in clustered networks than in unclustered
networks of the same degree distribution and degree correlations.

\section{Results}
\begin{figure}
\begin{center}
\includegraphics[width=0.8\columnwidth]{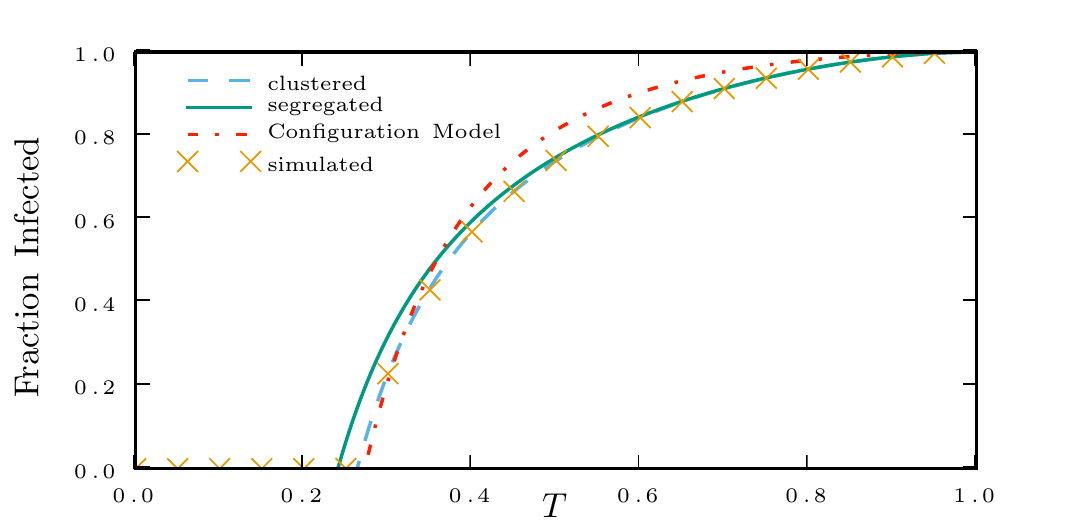}\\
\includegraphics[width=0.8\columnwidth]{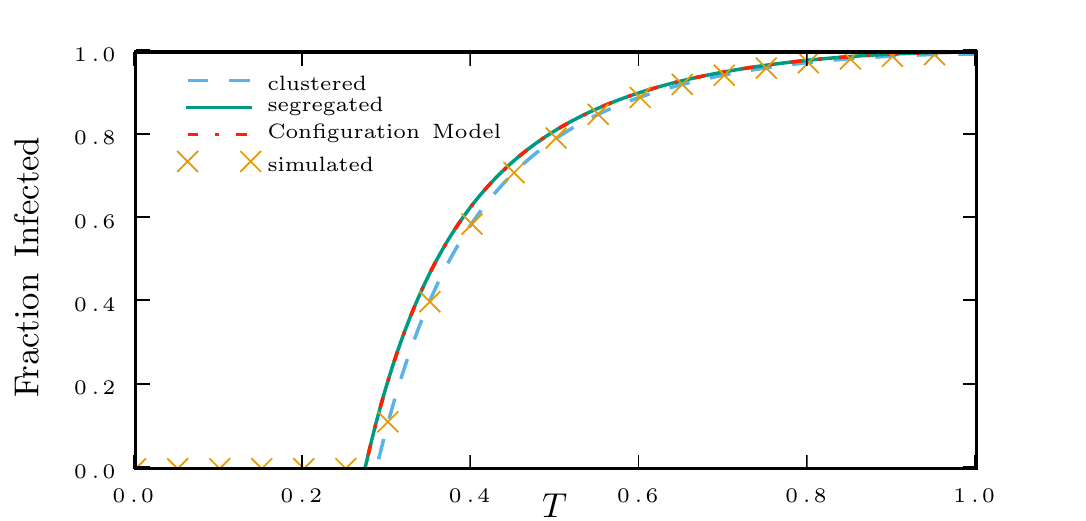}\\
\includegraphics[width=0.8\columnwidth]{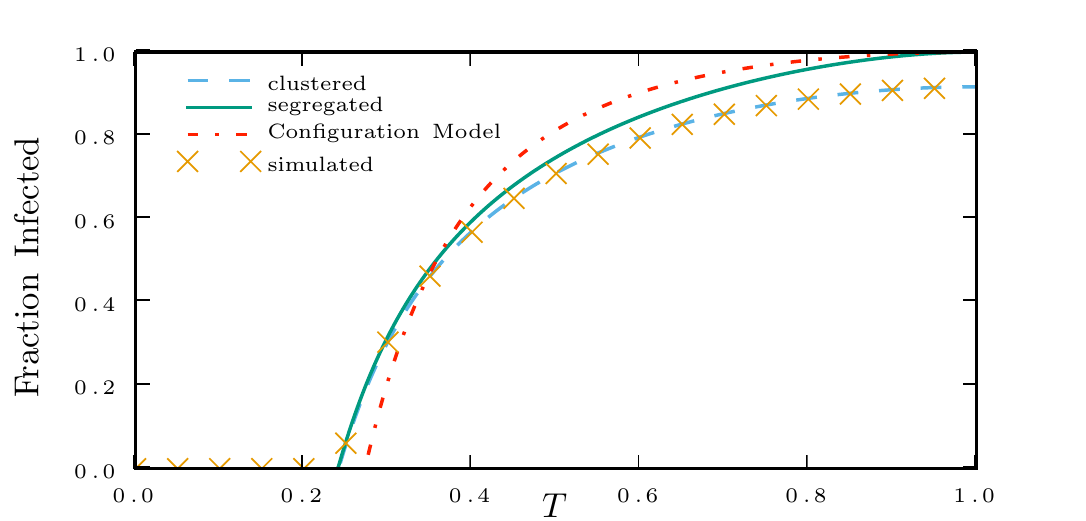}
\end{center}
\caption{A comparison of different network configurations.
  Assortative mixing reduces the epidemic threshold.  Clustering
  reduces epidemic size.}
\label{fig:comp}
\end{figure}

In figure~\ref{fig:comp} we consider outbreak spread on three
networks, all of which have the same degree distribution.  We compare
simulated epidemic sizes with predictions from the clustered
equations, the unclustered, segregated equations, and the equations
derived previously for configuration model
networks~\cite{newman:spread,miller:heterogeneity,kenah:second}.  

The nodes are equally distributed between degrees $2$, $4$, and $6$.
In each network the clustering is distributed differently.  In the
first, $p(0,3)=1/3$, \ $p(2,1) = 1/3$, and $p(2,0)=1/3$.  That is
those nodes with degree $6$ are only in triangles, nodes of degree $4$
have half of their edges in triangles and independent edges, and nodes
of degree $2$ have just independent edges.  High degree nodes tend to
be clustered and contact other high degree nodes.  The tendancy to
contact other high degree nodes reduces the epidemic threshold, but
the clustering raises the threshold.

In the second network, we take $p(2,0)=1/6$, \ $p(0,1)=1/6$, \
$p(2,1)=1/3$, \ $p(4,1)=1/6$, and $p(0,3)=1/6$.  This yields identical
distribution of neighbor degrees for nodes reached by either a
triangle or an independent edge.  The unclustered, segregated
equations yield the same result as the configuration model equations.
The clustered calculations have smaller epidemics.

The third network is an inversion of the first.  Nodes with
high degree have independent edges while nodes with low degree are
clustered.  We take $p(6,0)=1/3$, \ $p(2,1) = 1/3$, and $p(0,1)=1/3$.
Again the assortativity reduces the epidemic threshold while
clustering reduces the epidemic size.   In this particular case, it is
the preference for high degree nodes (which are unclustered) to
contact one another that leads to the reduction in epidemic threshold,
and so it is clear that the effect is due to assortative mixing, not clustering.

\section{Discussion}
We have introduced a new model of clustered networks on which we
study percolation and epidemics.  This model allows us to make a
number of analytic prediction because the edges of the network can be
partitioned into sets which are independent of one another
(independent edges or triangles).

We have shown that these networks can have a lower epidemic threshold
than Configuration Model networks with the same degree distribution.
However, this is not a consequence of clustering, but rather a
consequence of assortative mixing.  The clustering of the network
can be proven to raise the epidemic threshold and reduce the epidemic
size from networks with the same degree correlations, but without
clustering.

\section*{Acknowledgments}
This work was supported by the RAPIDD program of the Science \&
Technology Directorate, Department of Homeland Security and the
Fogarty International Center, National Institutes of Health.

\end{document}